\documentclass[twocolumn,aps,showpacs,prl,superscriptaddress]{revtex4}
\usepackage{amssymb}

\usepackage{epsfig}
\usepackage[english]{babel}
\usepackage{latexsym}
\usepackage{graphics}
\usepackage{subfigure}
\usepackage{epsfig}
\usepackage{graphicx}
\usepackage{dcolumn}
\usepackage{amsmath}
\usepackage{CJK}

\newcommand{\be}{\begin{equation}}
\newcommand{\ee}{\end{equation}}
\newcommand{\bey}{\begin{eqnarray}}
\newcommand{\eey}{\end{eqnarray}}
\newcommand{\bw}{\begin{widetext}}
\newcommand{\ew}{\end{widetext}}

\begin{document}
\title{Sensitive frequency-dependence of the carrier-envelope phase effect on bound-bound transition: an interference perspective}

\begin{CJK*}{UTF8}{gbsn}
\author{Dian Peng(彭典)}
\affiliation{Laboratory of Optical Physics, Beijing National Laboratory for Condensed Matter Physics,
Institute of Physics, Chinese Academy of Sciences, Beijing 100190, China}
\author{Biao Wu(吴飙)}
\affiliation{International Center for Quantum Materials, Peking
University, Beijing 100871, China}
\author{Panming Fu(傅盘铭)}
\affiliation{Laboratory of Optical Physics, Beijing National Laboratory for Condensed Matter Physics,
Institute of Physics, Chinese Academy of Sciences, Beijing 100190, China}
\author{Bingbing Wang(王兵兵)}
\email{wbb@aphy.iphy.ac.cn}
\affiliation{Laboratory of Optical Physics, Beijing National Laboratory for Condensed Matter Physics,
Institute of Physics, Chinese Academy of Sciences, Beijing 100190, China}
\author{Jiangbin Gong(龚江滨)}
\affiliation{Department of Physics and Center of
Computational Science and Engineering, National University of
Singapore, 117542, Singapore}
\author{Zong-Chao Yan(严宗朝)}
\affiliation{Department of Physics, University of New Brunswick, P. O. Box
4400, Fredericton, New Brunswick, Canada E3B 5A3}

\begin{abstract}
We investigate numerically with Hylleraas coordinates the
frequency dependence of the carrier-envelope phase (CEP) effect on
bound-bound transitions of helium induced by an ultrashort
laser pulse of few cycles. We find that the CEP effect is
very sensitive to the carrier frequency of the laser pulse,
occurring regularly even at far-off resonance frequencies. By
analyzing a two-level model, we find that the CEP effect can be
attributed to the quantum interference between neighboring
multi-photon transition pathways, which is made possible by the
broadened spectrum of the ultrashort laser pulse.  A general
picture is developed along this line to understand the sensitivity
of the CEP effect to laser's carrier frequency. Multi-level
influence on the CEP effect is also discussed.

\end{abstract}

\maketitle
\end{CJK*}
\par\noindent

\section{introduction}

For an ultrashort laser pulse that lasts for only a few cycles, its
carrier-envelope phase (CEP) can dramatically affect the yield of
matter-laser interaction~\cite{krausz}, leading to CEP dependence
of electron ionization~\cite{Paulus, 1, becker, li, lin} and
harmonic-photon emission~\cite{bohan, Krausz1, wang, krausz2}.
Recently, with the rapid development of the laser technology, CEP
has become a new way to control the dynamic process of
matter-laser interaction~\cite{krausz2}.
It has been demonstrated
that the CEP effect for an intense laser pulse can be measured by
comparing the photoelectron yields in two opposite directions
along the laser's electric field ~\cite{Paulus}.

More recently, the CEP effect on bound-bound transition of an atom
has been investigated theoretically~\cite{nakajima,Esry,
nakajima2} and observed experimentally~\cite{scully}. Roudnev and
Esry have presented a general framework for understanding the CEP
effect using the Floquet theory~\cite{Esry}. Li \emph{et al.}
have demonstrated that an experimentally observed CEP effect can be
attributed to the interference between one- and three-photon
transition pathways~\cite{scully}. The study by Nakajima and
Watanabe suggests that the CEP effect can occur as the laser's carrier frequency is far
off-resonance~\cite{nakajima}.

In this paper, we use Hylleraas coordinates to study
numerically how the CEP effect on bound-bound transitions of helium
changes as a function of the carrier frequency of an ultrashort laser.
Our numerical results show that the CEP effect depends
sensitively on the carrier frequency even when it is far off-resonance.
The essential physics implied in these numerical results
can be well revealed by a two-level model.  For a two-level system,
when the pulse duration is long, quantum transitions peak
at well-separated multi-photon resonant frequencies. As the
pulse duration decreases to less than three laser cycles, for example,
the widths of such transition peaks get significantly broadened,
and eventually two broadened neighboring transition peaks
can cross with each other. As a result, two different multi-photon
transitions  can both contribute significantly to the total transition amplitude
and hence interfere with each other. We show  that a large CEP
effect occurs exactly at these crossings and can hence be
understood as a quantum interference effect~\cite{brumer}. A general
and simple picture developed along this line enables us
to clearly answer
the following questions: (1) Why is the CEP effect on bound-bound
transitions  sensitive to a carrier frequency that is far-off
resonance? (2) Which multi-photon transition pathways
can interfere and lead to the CEP effect?

\section{numerical method towards the CEP effect in Helium}
\label {sec2}

Atomic units are used throughout unless specified otherwise.
To study a photoexcitation process of a ground state helium atom
in a linearly polarized ultrashort laser pulse, we solve a
time-dependent Schr\"{o}dinger equation
$i\frac{\partial}{\partial
t}|\Psi(t)\rangle=[\textbf{H}_0+\textbf{H}_1(t)]|\Psi(t)\rangle$. The field-free part of the
Hamiltonian reads
\begin{equation}\label{eq:Ham}
\textbf{H}_0=-\frac{1}{2}\nabla^2_1-\frac{1}{2}\nabla^2_2-\frac{2}{r_1}-\frac{2}{r_2}+
\frac{1}{|{\textbf{r}_1}-{\textbf{r}_2}|}\;,
\end{equation}
where ${\textbf{r}}_1$ and ${\textbf{r}}_2$ are the coordinates of
the two electrons measured from the nucleus located at the origin.
The light-atom interaction part of the Hamiltonian is
\begin{equation}
\textbf{H}_1=({\textbf{r}}_1+{\textbf{r}}_2)\cdot\textbf{e}(t)=
-({\textbf{r}}_1+{\textbf{r}}_2)\cdot\partial\textbf{A}(t)/\partial t\;,
\end{equation}
where the vector potential of the field is given by~\cite{x, becker}
\begin{eqnarray}
\textbf{A}(t)=\hat{\mathbf{\varepsilon}}A_0\exp(-\alpha^2t^2)\sin(\omega
t+\phi)/\omega,
\end{eqnarray}
with $\hat{\varepsilon}$ being the linear polarization
vector, $\omega$ the carrier frequency, and $\phi$ the CEP parameter.
The full width at half maximum
(FWHM) of the pulse duration is $\tau=2\sqrt{\ln2}/\alpha$.

The wave function of helium $|\Psi(t)\rangle$ is expanded in terms of the
field-free eigenvectors $|\psi_n\rangle$:
$|\Psi(t)\rangle=\sum_na_n(t)e^{-iE_nt}|\psi_n\rangle$,
where $\textbf{H}_0|\psi_n\rangle=E_n|\psi_n\rangle$. Then
we numerically solve the following set of equations for the amplitudes
$a_n(t)$
\begin{equation}\label{eq:ode}
i\frac{d}{dt}a_n(t)=\sum_{m}a_m(t)e^{-iE_{mn}t}H^I_{nm}(t)\;,
\end{equation}
where $E_{mn}=E_m-E_n$, and
$H^I_{nm}(t)=\langle\psi_n|\textbf{H}_1|\psi_m\rangle$. Initially,
the helium is in the ground state. The basis set is constructed in
Hylleraas coordinates~\cite{yan}
\begin{eqnarray*}
  |\psi_n\rangle=\sum_mc_{nm}|\phi_m\rangle,
\end{eqnarray*}
where
\begin{eqnarray*}
|\phi_m\rangle=r_1^ir_2^jr_{12}^ke^{-\alpha r_1-\beta
r_2}\mathcal{Y}_{l_1l_2}^{LM}(\textbf{r}_1,\textbf{r}_2).
\end{eqnarray*}
In the above, $\mathcal{Y}_{l_1l_2}^{LM}(\textbf{r}_1,\textbf{r}_2)$ is the
vector coupled product of spherical harmonics for the two
electrons. In order to obtain convergent results, 874 Hylleraas functions
are used  with the total angular momentum quantum number $L$ ranging
from 0 to 4.

Our numerical results show that the population of an excited state of helium
can strongly depend on the CEP parameter $\phi$ at certain carrier
frequencies. To quantify this CEP effect, we
introduce the parameter~\cite{nakajima}
\begin{equation}\label{eq:cepQ}
{\cal M}=\frac{P(\phi_{\rm max})-P(\phi_{\rm min})}{(P(\phi_{\rm max})+P(\phi_{\rm min}))/2}\,,
\end{equation}
where $P(\phi_{\rm max})$ and $P(\phi_{\rm min})$ are, respectively,
the maximum and minimum  populations for a given excited state.  A large value of
${\cal M}$ means a strong CEP effect. We numerically
calculate ${\cal M}$ for the $2^1P$ and $3^1D$ states.
In our computation, the FWHM pulse
duration is one laser cycle and the peak intensity of the laser is
$10^{13}$ W/cm$^2$.

\begin{figure}[!htb]
\begin{center}
 \includegraphics[width=0.45\textwidth]{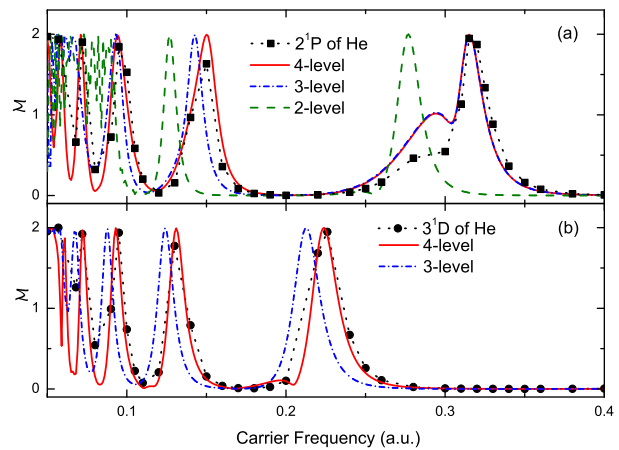}
\end{center}
\caption{(color online) The CEP parameter ${\cal M}$ [see Eq.(\ref{eq:cepQ})] versus
laser's carrier frequency for the 2$^1P$ state (a) and the 3$^1D$
state (b) of helium. Solid squares and solid circles are numerical
results, (green) dashed line is for a two-level model, (blue)
dot-dashed lines are for a three-level model, and (red) solid lines
are for a four-level model.} \label{fig:m1}
\end{figure}

Figure~\ref{fig:m1} presents how ${\cal M}$ changes with the laser
carrier frequency $\omega$. Results for the $2^1P$ state (squares)
are shown in panel (a) and those for the $3^1D$ state (circles)
are shown in panel (b). It is clear from the figure that there are
many carrier frequency windows in which ${\cal M}$ peaks at about
its maximum value of 2, indicating a strong CEP effect on the
transition probabilities from the ground state to the two excited
states. These numerical details qualitatively agree with those
reported in Ref.~\cite{scully}, where the authors changed the
energy difference between two bound states instead.

One important feature in Fig. \ref{fig:m1} is that the CEP effect is
very sensitive to the
carrier frequency even it is  far off-resonance as compared with the
energy difference between the two bound states. This implies that
 far off-resonance is not a sufficient condition for  a large CEP effect. In
the following, we attempt to answer the following question: what
is the physics underlying these narrow frequency windows where  the
parameter ${\cal M}$ peaks?

We have compared our full numerical results in
Fig.~\ref{fig:m1}~(a) to the results of a two-level model (to be
elaborated below). As clearly shown in the figure, except the peak
positions are shifted towards lower frequencies, the two-level
results (the (green) dashed curve in Fig.~\ref{fig:m1}~(a)) can
embody the main features in the full numerical results for
$\omega > 0.1$. This indicates that a simple two-level model is
sufficient to reflect the essential physics behind these peaks. In
the following, we first study a two-level model, and then
investigate multi-level influence on the CEP effect.

\section{two-level model and interference between neighboring transition pathways}
Consider a two-level system in a pulsed laser field. The
ground and the first excited states of helium are chosen as the two
levels, denoted as $|0\rangle$ and $|1\rangle$. According to
Eq.~(\ref{eq:ode}), the amplitudes of these two states obey the
following equations
\begin{eqnarray}\label{eq:ode2}
\dot{a}_0(t)&=&i\mu_{01}f_{01}a_1(t)\,,\nonumber\\
\dot{a}_1(t)&=&i\mu_{10}f_{10}a_0(t)\,,
\end{eqnarray}
where $\mu_{jk}= \mu_{kj}$ represents the transition dipole moment
between two quantum states $|j\rangle$ and $|k\rangle$ and
$f_{jk}(t)\equiv e(t)\exp(i\Delta E_{jk}t)$ with $e(t)$ being
laser's electric field and $\Delta E_{jk}$ the energy difference
between $|j\rangle$ and $|k\rangle$. 

If the system is initially in the ground state,
a formal solution of $a_1$, after the laser pulse has passed, can be
written as
\begin{eqnarray}\label{eq:a1}
a_1&=&i\mu_{01}\int_{-\infty}^{\infty}dtf_{10}(t)\nonumber \\
&+&(i\mu_{01})^3\int_{-\infty}^{\infty}\int_{-\infty}^{t}
         \int_{-\infty}^{t_1}f_{10}(t)f_{01}(t_1)f_{10}(t_2)dtdt_1dt_2\nonumber \\
 &+&   \cdots\nonumber \\*
&\equiv &\sum_{n=0}^\infty T^{(2n+1)} ,
\end{eqnarray}
where $T^{(2n+1)}$ represents  the
$(2n+1)$-photon \textit{quantum transition amplitude}.
For example, $T^{(1)}$ is for one-photon
transition amplitude and  $T^{(3)}$ is for three-photon
transition amplitude.

\begin{figure}[!htb]
\includegraphics[width=0.35\textwidth]{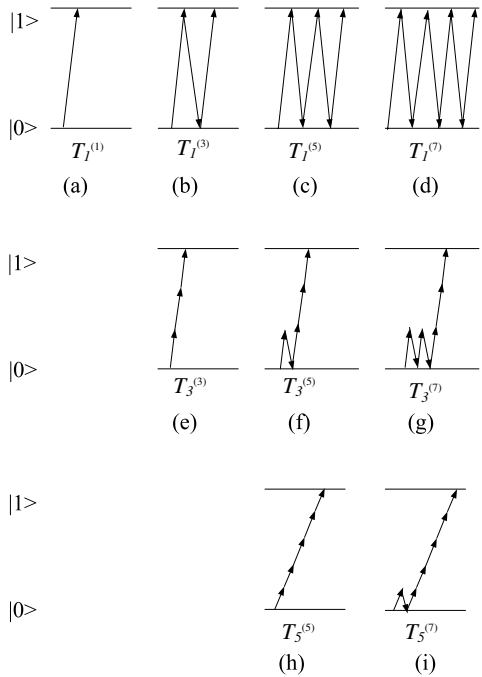}
\caption{Quantum transition pathways between two bound states
of an atom. For the pathways in the same column, they involve
the same number of photons, e.g., the third column is for
five-photon pathways. For the pathways in the same row,
they share the same resonance frequency, e.g.,
the pathways in the second row have the resonance frequency
$\Delta E_{10}/3$, where
$\triangle E_{10}$ is the energy spacing between the two bound states.}\label{fig:level1}
\end{figure}

To clearly understand the physics behind each amplitude $T^{(2n+1)}$,
we first consider the limit of long pulse duration, i.e., $\alpha\rightarrow 0$,
which is easier to deal with. In this limit, after neglecting the
negative frequency components~\cite{scully},
the $(2n+1)$-photon transition amplitude $T^{(2n+1)}$ can be
decomposed into a sum of $(n+1)$ terms, i.e.,
\begin{equation}\label{eq:tp}
T^{(2n+1)}\approx \sum_{j=0}^n T_{2j+1}^{(2n+1)},
\end{equation}
where
\begin{equation}\label{eq:delta}
T_{2j+1}^{(2n+1)}\propto \delta\left[\Delta E_{10} -
(2j+1)\omega\right]\exp[-i(2j+1)\phi]\,.
\end{equation}
This implies that,
physically, each term $T_{2j+1}^{(2n+1)}$ can be associated with a
quantum transition pathway, which carries a phase $-(2j+1)\phi$
and contributes significantly to the overall transition amplitude
at resonant frequency $\omega_j=\Delta E_{10}/(2j+1)$.  These
quantum transition pathways are depicted schematically in
Figure~\ref{fig:level1}, where the $m$th column shows all the
pathways involving $(2m-1)$ photons and the $k$th row includes all
the pathways that have a resonant frequency at
$\omega_{k-1}=\Delta E_{10}/(2k-1)$ with $m\leq 4$ and $k\leq 3$.

With the decomposition in Eq.(\ref{eq:tp}), it is clear that the
overall transition amplitude $a_1$ is significantly different from
zero only at resonance frequencies. At the same time,  we notice
that all the pathways at a given resonance frequency $\omega_j$
carry the same phase $-(2j+1)\phi$. Therefore, the CEP $\phi$ does
not affect the magnitude of $a_1$. This clearly explains why there
is no CEP effect for a long-pulsed laser.

\begin{figure}[!htb]
\includegraphics[width=0.45\textwidth]{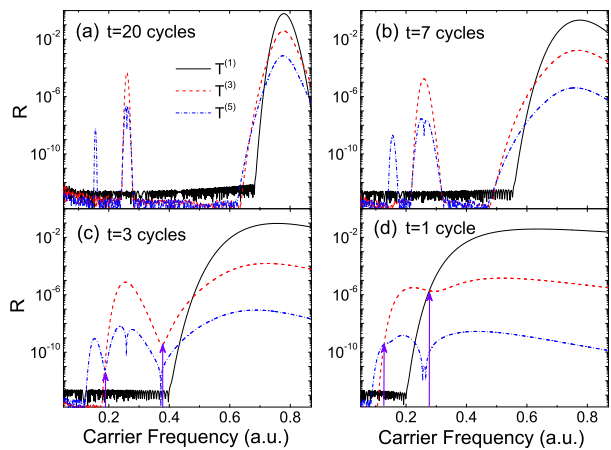}
\caption{The magnitude of $T^{(2n+1)}$ with $n=0$ (solid curves),
$n=1$ (dashed curves), and $n=2$ (dash-dotted curves) for pulse
duration (a) $\tau =20$, (b) $\tau=7$, (c) $\tau=3$, and (d)
$\tau=1$ laser cycle(s). The arrows in panel (c) indicate the
interference between $T_3^{(3)}$ and
$T_1^{(3)}$ (right) and $T_3^{(3)}$ and
$T_5^{(5)}$ (left); the arrows in panel (d) indicate the
interference between $T_1^{(1)}$ and
$T_3^{(3)}$ (right) and $T_3^{(3)}$ and
$T_5^{(5)}$ (left).}\label{fig:tpd}
\end{figure}

The situation becomes very different as the laser pulse becomes shorter.
It is reasonable to assume that the decomposition in Eq.(\ref{eq:tp}) still
holds even for short laser pulse. However, as the pulse becomes shorter,
$T_{2j+1}^{(2n+1)}$ is no longer proportional to a delta function as in Eq.(\ref{eq:delta}),
but becomes a function of the laser's frequency which peaks at $\omega_j=\Delta E_{10}/(2j+1)$. Moreover, the width of the peak
associated with each transition path $T_{2j+1}^{(2n+1)}$
broadens as the pulse duration gets shorter.
Consequently, when the pulse is very short, e.g.,
lasting only for a few cycles, the peaks can become so wide that
the peaks for pathways $T_{2j+1}^{(2n+1)}$ with different $j$'s can
cross with each other at a certain far off-resonance frequency.
As a result, the overall transition amplitude $a_1$ can be regarded as
an interference between the two pathways, if it is dominated by
this pair of pathways at the crossing point.
At the same time, we notice that each pathway $T_{2j+1}^{(2n+1)}$ carries
the phase $-(2j+1)\phi$, indicating that the relative phase between this
pair of pathways depend on the CEP $\phi$. These facts mean that
the overall transition amplitude $a_1$ depends on $\phi$  at
the crossing point and is strongly affected by the  CEP.

The above analysis is confirmed by our detailed numerical calculations
shown in Fig.~\ref{fig:tpd}, where we analyze the amplitude
$R=|T^{(2n+1)}|$ as a function of the laser's carrier frequency for
different pulse durations while the CEP is fixed at $\phi=0$. As
clearly demonstrated in Fig.~\ref{fig:tpd}(a), for a laser pulse
of 20 cycles, the peaks at resonance frequencies $\omega_0$,
$\omega_1$, and $\omega_2$ are narrow and well separated.
Particularly, in consistent with the above picture for the long pulse
limit, one finds that the magnitude of $T^{(1)}$ has one peak
located at $\omega_0$, indicating the contribution from
$T_1^{(1)}$; whereas the magnitude of $T^{(3)}$ has two
peaks located at $\omega_0$ and $\omega_1$, indicating the
respective contributions from $T_1^{(3)}$
and $T_3^{(3)}$. Finally the magnitude of $T^{(5)}$ has
three peaks located at $\omega_0$, $\omega_1$, and $\omega_2$, due
to $T_1^{(5)}$, $T_3^{(5)}$,
and $T_5^{(5)}$, respectively. As the pulse is shortened
to seven cycles, there is no essential change in the overall
features: only each peak becomes wider, as demonstrated in
Fig.~\ref{fig:tpd}(b), which shows that the decomposition for short
pulses in  Eq.(\ref{eq:tp}) is well justified.

It is a very different situation when the pulse is shortened to 3
cycles. In this case, the peaks become so broad that they begin to
overlap and cross into each other. As indicated by arrows in
Fig.~\ref{fig:tpd} (c), there are two crossings. One occurs  at
$\omega=0.380$, where the peak associated with
$T_1^{(3)}$ crosses with the peak associated with
$T_3^{(3)}$. The other crossing happens at
$\omega=0.185$, where the peak associated with
$T_3^{(3)}$ crosses with the peak associated with
$T_5^{(5)}$. At such crossing points, the two dominant
pathways have the same magnitude, which is essential for effective
interference and strong CEP effect. Finally, when the pulse
duration is decreased to one laser cycle in Fig.~\ref{fig:tpd}~(d),
the contributions from the pathways $T_1^{(1)}$ and
$T_3^{(3)}$ cross with each other at $\omega=0.277$ and
the contributions from $T_3^{(3)}$ and $T_5^{(5)}$ cross at
$\omega=0.127$. Compared with the two-level results in
Fig.~\ref{fig:m1}, it is seen that the CEP effect occurs exactly
at the frequencies where the crossings between the broadened
peaks occur.

\begin{figure}[!htb]
\includegraphics[width=0.45\textwidth]{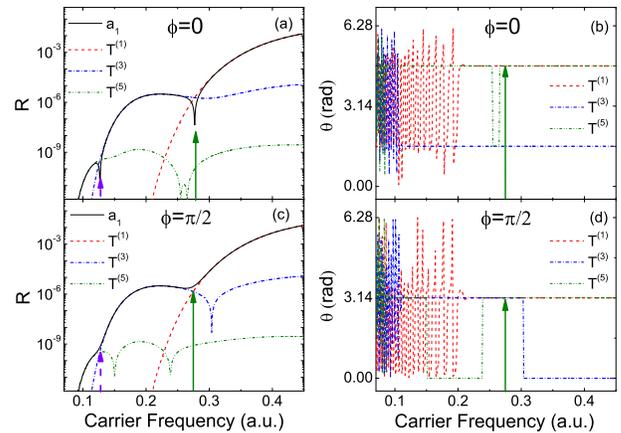}
\caption{Multi-photon transition contributions to the excited
state amplitudes $T^{(2n+1)}$ with $n=0, 1, 2$, as a function of
laser's carrier frequency.  The CEP parameter is given by $\phi=0$
in panels (a) and (b), and $\phi=\pi/2$ in panels (c) and
(d).}\label{fig:tp}
\end{figure}

To see how the CEP affects the transition amplitude,
 we next investigate both the magnitudes and the phases of
$T^{(1)}$, $T^{(3)}$, and $T^{(5)}$ by setting
$T^{(2n+1)}=R_{T^{(2n+1)}}\exp{(i\theta_{T^{(2n+1)}})}$
 with the pulse duration fixed at one
laser cycle.  The results are shown in Fig.~\ref{fig:tp}(a) and
(b) for $\phi=0$ and Fig.~\ref{fig:tp}(c) and (d) for
$\phi=\pi/2$.  Arrows in Fig.~\ref{fig:tp} (a) indicate the two
frequencies $\omega=0.277$ (solid arrows) and $\omega=0.127$
(dashed arrows) at which the CEP effect occurs. At $\omega=0.277$,
where the curves for $R_{T^{(1)}}$ and $R_{T^{(3)}}$ cross with each other, we see
from Fig.~\ref{fig:tp} (b) and (d) that  the phase difference
$\theta_{T^{(1)}}-\theta_{T^{(3)}}$ is around $\pi$ at
$\phi=0$ and around zero at $\phi=\pi/2$. This means that when
$\phi=\pi/2$, the contributions to the total transition amplitude from
$T^{(1)}$ and ${T^{(3)}}$ constructively interfere, whereas at
$\phi=0$ they destructively interfere, leading to a
significant CEP effect at $\omega=0.277$.  Therefore, the CEP
effect is the result of  the $\phi$-dependent interference between
the one-photon contribution $T^{(1)}$ and the three-photon
contribution $T^{(3)}$.  Similarly, the strong CEP
effect at $\omega=0.127$ can be attributed to the $\phi$-dependent
interference between the three-photon amplitude $T^{(3)}$ and the
five-photon amplitude $T^{(5)}$.

Based on this simple picture, it is apparent that the CEP
effect often appears at a carrier frequency that is \emph{off
resonance}: the contributions to the total transition amplitude
from two different pathways with different $j$'s can be comparable
only at a frequency that is \emph{not} equal to the resonance
frequency $\omega_j$.

In principle, if the pulse duration is short enough, the transition
amplitudes, which are contributed by any two neighboring pathways associated
with different resonance frequencies $\omega_j$'s,  can
interfere with each other.  As such, one might expect the curve of
$\cal{M}$ vs $\omega$ to have an infinite number of peaks.
However, since the transition amplitude associated with a
multi-photon pathway decreases rapidly with the number of
photons involved, the actual number of clear CEP peaks will be limited by
the highest order of multi-photon transitions that can have a
significant transition amplitude. For example, in our two-level
model with the laser intensity adopted, the highest multi-photon
order is five, thus yielding only two peaks in the curve of
$\cal{M}$ vs $\omega$; these two peaks arise from the interference between one-
and three-photon pathways, as well as between three- and five-photon
pathways. If the laser intensity is further increased, then more CEP peaks can be expected.
Scully \emph{et al.}~\cite{scully} presented an
example of the CEP effect caused by the interference
between the one- and three-photon pathways, which is just one of many
possible peaks in our general picture.

\begin{figure}[!htb]
\includegraphics[width=0.45\textwidth]{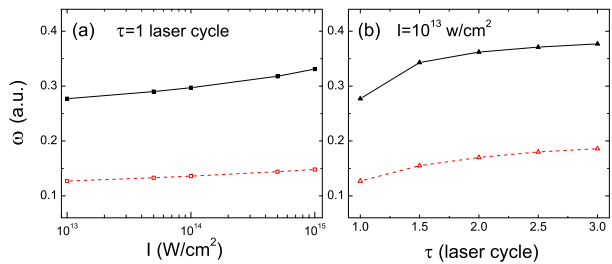}
\caption{The frequency versus the laser intensity (a) and the pulse duration (b) at
which the CEP effect occurs in the two-level model.}\label{fig:a1-d-i}
\end{figure}

We next describe some computational details regarding how the
laser intensity and the laser pulse duration change the CEP effect.
Figure~\ref{fig:a1-d-i}(a) depicts how the two CEP frequencies
(i.e., the carrier frequencies at which the ${\cal M}$ can
reach a large value) vary with the laser intensity, where the pulse
duration is fixed at $\tau=1$ laser cycle. Figure~\ref{fig:a1-d-i}(b)
shows the impact of the pulse duration, with the laser intensity
fixed at $10^{13}$~W/cm$^2$. It is seen from Fig.~\ref{fig:a1-d-i}(a)
that the CEP frequency increases slightly
as the intensity increases. This is because the magnitude of the
multi-photon transition contribution $T^{(2n+1)}$ is proportional
to $I^{(2n+1)/2}$, with $I$ being the laser intensity. As such,
the amplitudes of higher order multi-photon transitions increase
with the laser intensity much faster than those of lower order
ones, resulting in that the crossing of neighboring peaks
associated with different transition pathways occurs at higher
frequencies as the laser intensity increases.  Turning to the
pulse duration dependence shown in Fig.~\ref{fig:a1-d-i}(b), it is
observed that the CEP frequencies are also slightly blue shifted
as the pulse duration increases, indicating that the crossing
between neighboring peaks occurs at slightly higher frequencies as
the pulse duration increases. When the pulse duration is larger
than about four laser cycles, the CEP effect disappears for $I=
10^{13}$~W/cm$^2$, implying the vanishing of crossing between
any two neighboring multi-photon transition pathways.

\begin{figure}[htb]
\includegraphics[width=0.45\textwidth]{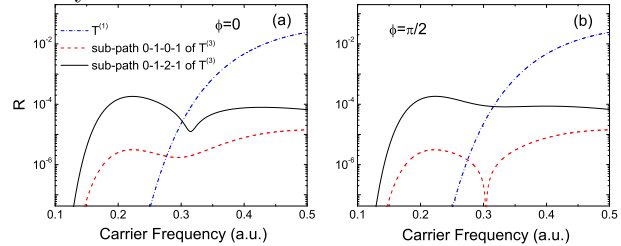}
\caption{The magnitude $R$ of $T^{(1)}$, as well as the
magnitudes of the transition amplitudes formed respectively by the sub-path
0-1-0-1 (dashed line) and by the sub-path 0-1-2-1 (solid line)
associated with $T^{(3)}$, as a function of the laser's carrier
frequency. The results are obtained using a three-level model. In
panel (a) the CEP parameter is given by $\phi=0$ and in panel (b)
the CEP parameter is given by $\phi=\pi/2$.
}\label{fig:3level-tp1}
\end{figure}

\section{multi-level effect }
For a real atom, laser-atom interaction induces transitions among
many states. As such, we need to consider the influence of
multi-level transitions on the final population of the state of
interest. As shown in Fig.~\ref{fig:m1}~(a), the results of the
two-level model disagree quantitatively with the full numerical
calculations: the two peaks at $\omega=0.277$ and $\omega=0.127$ appeared in the
two-level model are red-shifted as compared to the peaks at
$\omega=0.315$ and $\omega=0.15$ found in the full numerical
approach.  One is then motivated to examine a three-level model
that incorporates $|0\rangle$, $|1\rangle$, and $|2\rangle =3^1D$
states of helium. Interestingly, the first peak found in this 3-level
model agrees well with the full numerical calculations, but the other
peaks are still red-shifted as compared with those found in the
full numerical calculations. Finally, we include one more level, i.e.,
$|3\rangle =4^1F$, in the dynamics and hence obtain a four-level
model.  The results from such a four-level model are found to
agree well with the full numerical results for the frequency
region considered here.  Detailed comparisons between these
models are presented in Fig.~\ref{fig:m1}.

The above-mentioned quantitative differences between various
models can be explained by analyzing the transition pathways in
multi-level situations. Consider Eq.~(\ref{eq:a1}) again. Now the
term $T^{(3)}$ for the three-level model includes two integrals:
\begin{widetext}
\begin{eqnarray}\label{eq:tp3-3}
T^{(3)}&=&(i\mu_{01})^3\int_{-\infty}^{\infty}\int_{-\infty}^{t}
         \int_{-\infty}^{t_1}f_{10}(t)f_{01}(t_1)f_{10}(t_2)dtdt_1dt_2\nonumber \\*
         &&+(i\mu_{01})(i\mu_{12})^2\int_{-\infty}^{\infty}\int_{-\infty}^{t}
         \int_{-\infty}^{t_1}f_{12}(t)f_{21}(t_1)f_{10}(t_2)dtdt_1dt_2,
\end{eqnarray}
\end{widetext}
where the two integrals correspond to the following two sub-paths:
$|0\rangle\rightarrow|1\rangle\rightarrow|0\rangle\rightarrow|1\rangle$
(0-1-0-1) and
$|0\rangle\rightarrow|1\rangle\rightarrow|2\rangle\rightarrow|1\rangle$
(0-1-2-1). Clearly both sub-paths can contribute
significantly to the total quantum transition amplitude from state
$|0\rangle$ to state $|1\rangle$. Furthermore,  the transition
dipole moment $\mu_{12}$ between states $|1\rangle$ and
$|2\rangle$ is larger than the transition dipole moment $\mu_{01}$
between states $|0\rangle$ and $|1\rangle$ (see Table I). This
indicates that the transition amplitude from $|1\rangle$ to
$|2\rangle$ is more substantial than that from $|0\rangle$ to
$|1\rangle$.  It can be estimated that the contribution by the
sub-path 0-1-2-1 of $T^{(3)}$ is about one order of magnitude
larger than that by the sub-path 0-1-0-1.  A detailed numerical
comparison between them is shown in Fig.~\ref{fig:3level-tp1}.
Therefore the role of the sub-path 0-1-0-1 in the previous
two-level model should be replaced by the sub-path of 0-1-2-1 in
the present three-level model, thus explaining the quantitative
difference between a two-level model and a three-level model.

Similarly, for the interference between the pathways of $T^{(3)}$
and $T^{(5)}$ that accounts for the second peak at $\omega=0.15$
in Fig.~\ref{fig:m1}(a), the contribution of the sub-path
0-1-2-3-2-1 to the magnitude of $T^{(5)}$ is larger than the
contributions of the sub-paths 0-1-2-1-2-1, 0-1-2-1-0-1, and
0-1-0-1-2-1. This can be interpreted by a four-level model. In
addition, as for interference between multi-photon transitions of
even higher orders, the results of the four-level model also agree
well with the full numerical calculations (Fig.1(a)). This is because
that the populations of higher states are so small under the present laser
conditions that the additional sub-paths, including these higher
states, are not important regardless of their larger transition
dipole moments.

\begin{table}[htb]
\caption{Transition dipole moment of helium in atomic units.  As
discussed in the text, the big differences in the transition
dipole moments are an important factor when explaining the
quantitative differences between two-level, three-level, and
four-level models. \\ }
\begin{tabular}{|c|c|c|}
\hline
$\mu_{01}$& $\mu_{12}$ &$\mu_{23}$ \\
\hline
0.4208 & 2.499 & 5.175 \\
\hline
\end{tabular}\\[0.5ex]
\end{table}

Finally, we return to Fig.~\ref{fig:m1}(b), where $\cal{M}$
versus the laser's carrier frequency is shown for the $3^1D$
state of helium. Similar to the case of $2^1P$, the peaks in
Fig.~\ref{fig:m1}(b) can be associated with the interference between two
neighboring transition paths at $\omega=\Delta E/(2n)$,
where $n=1,2,\ldots$, and $\Delta E$ is the energy spacing
between the ground state and the $3^1D$ state.  Our general
picture above can equally be applied to analyze this case and the details will
not be repeated here.

\section{Summary}
\label {sec5}

In summary, we have investigated numerically the frequency
dependence of the carrier-envelope phase effect on bound-bound
transitions of helium in an ultrashort laser pulse. It has been found
that the CEP effect can occur regularly even at frequencies which
are far off-resonance. To explain this numerical finding, we have
examined a two-level model and developed a general and simple
picture, where the total transition amplitude can be decomposed
into different transition pathways. All these transition pathways
can be  characterized by two indices $n$ and $j$, with $n$ being
the number of photons involved and $j$ indicating where the
resonance frequency is located. For a long laser pulse, each
pathway is associated with a narrow peak at the resonance frequency
$\omega_j$ and the pathways at the same resonance frequency have
the same dependence on the CEP. As a result, no interference can
occur between different pathways and there is no CEP effect. In
contrast, for an ultrashort pulse, the peaks for pathways at
different resonance frequencies can be broadened to cross with each
other. Therefore, interference between different pathways can
happen and lead to strong CEP effect. This general picture is
valid for a wide range of laser intensities as long as the
perturbation method is applicable, and can be generalized to
multi-level models.

\begin{acknowledgments}
This research was supported by the National Natural Science
Foundation of China under Grant No. 60778009, No. 11074296, and
No. 10825417, and the 973 Research Project No. 2006CB806003.
ZCY was supported by NSERC of Canada and by the Canadian
computing facilities ACEnet, SHARCnet, and WestGrid.
\end{acknowledgments}

\end{document}